# Overlooked transportation anisotropies in *d*-band correlated rare-earth perovskite nickelates

*Jikun Chen[1]\*, Haiyang Hu[1], Fanqi Meng[2], Takeaki Yajima[3], Lixia Yang[2], Binghui Ge[2,4]\*, Xinyou Ke[5], Jiaou Wang[6], Yong Jiang[1], Nuofu Chen[7]*

[1]Beijing Advanced Innovation Center for Materials Genome Engineering, School of Materials Science and Engineering, University of Science and Technology Beijing, Beijing 100083, China

[2]Beijing National Laboratory for Condensed Matter Physics, Institute of Physics, Chinese Academy of Sciences, 100190 Beijing, China

[3]School of Engineering, The University of Tokyo, 2-11-16 Yayoi, Bunkyo-ku, Tokyo 113-0032, Japan

[4]Institutes of Physical Science and Information Technology, Anhui University, Heifei, Anhui 230601, China

[5]Key Laboratory of Structure and Functional Regulation of Hybrid Materials (Anhui University), Ministry of Education, Heifei, Anhui 230601, China

[6]Beijing Synchrotron Radiation Facility, Institute of High Energy Physics, Chinese Academy of Sciences, Beijing 100049, China

[7]School of Renewable Energy, North China Electric Power University, Beijing 102206, China

Correspondence: Prof. Jikun Chen (jikunchen@ustb.edu.cn) and Prof. Binghui Ge (bhge@ahu.edu.cn)

.

**Abstract**

Anisotropies in electronic transportations conventionally originate from the nature of low symmetries in crystal structures, and were not anticipated for perovskite oxides, the crystal asymmetricity of which is far below, e.g. van der Waals or topological crystal. Beyond conventional expectations, herein we demonstrate pronounced anisotropies in the inter-band coulomb repulsion dominated electronic transportation behaviors under low-dimensional confinement for the perovskite family of rare-earth nickelates (*Re*NiO$_3$). From one aspect, imparting bi-axial interfacial strains upon various lattice planes results in extrinsic anisotropies in the abrupt orbital transitions of *Re*NiO$_3$, and their metal to insulator transition behaviors that elevates the transition temperature beyond the existing merit. From the other aspect, the in-plane orbital entropy associated to the in-plane symmetry of the NiO$_6$ octahedron within *Re*NiO$_3$ causes intrinsic anisotropies for the gradually orbital transition with temperature to further improve their thermistor transportation properties. The present work unveils the overlooked role of the electronic orbital directionality within low dimensional correlated perovskites that can trigger anisotropic transportation behaviors, in spite of their relatively symmetric crystal structures. Establishing anisotropic transportations integrating the electron correlation and quantum confinement effects will bring in a new freedom for achieving further improvement in transportation properties of multi-functional perovskite oxides.

Anisotropic material properties are usually observed in crystal structures at a low symmetry, such as graphene/CNT-based composites [1], layer-structured van der Waals crystals [2,3] and topological semiconductors [4-6]. In contrast to these layer-structured materials, perovskite structured oxides ($ABO_3$) exhibit a higher symmetry in crystal structures, and thereby the anisotropy for $ABO_3$ is not expected as a critical issue in their electrical transportations and applications. Nevertheless, distinguished features for typical transition metal perovskite oxides, beyond conventional semiconductors, are associated to their more complicated orbital configurations and unneglectable inter-band Coulomb interactions [7-10]. The resultant electron-electron and/or electron-phonon coupling within condensed matters enriches additional transportation characteristics, such as metal to insulator transitions (MIT) [7], high-$T_C$ superconductance [8], bad metal [9] and thermistor transports [10]. It is worthy to note that in contrast to conventional carrier conductions, the Coulomb coupling of the carrier transportation with either the orbital bands or phonons at an electron more localized manner should be directional even for more symmetric crystal structures [11-14]. It raises a fundamental open question within condensed matter physics: ***Whether there are any anisotropic transportation behaviors, either intrinsically exist or can be artificially established, within the electron correlated perovskite oxides?***

The rare-earth nickelates ($Re$NiO$_3$) is a representative material family of $d$-band correlation perovskite oxides, known for their extremely complex electronic phase diagram and widely adjustable MIT transportations [9-14]. Below their transition temperature ($T_{MIT}$), energy band gaps are splitted from hybridized Ni-3$d$ and O-2$p$ orbitals, and their orbital configurations are highly susceptible to the construction of NiO$_6$ octahedron. As a result, the electronic transportation is widely regulatable for the insulating phase of $Re$NiO$_3$, by simply regulating the rare-earth composition [10,11], temperature [10-13], strain-induced lattice distortions [11-14] or electronic polarizations [15]. It is worth noticing that carrier transportation within the insulating phase of $Re$NiO$_3$ follows a hopping principle along the head-to-head oriented O-2$p$ and Ni-3$d$ orbitals within the network of inter-linked NiO$_6$ octahedrons [11]. If such transport can be spatially confined within two or lower dimension, the anisotropy in the correlated transportation properties will be unveiled owing to the intrinsic anisotropy in the orbital configurations within a single crystalline $Re$NiO$_3$. This is expected to bring in a new freedom in regulating correlated transportation properties of $Re$NiO$_3$ thin films via controlling preferential orientations under the dimension confinement. As a result, the adjustability in achievable MIT or thermistor transportation properties of $Re$NiO$_3$ will be broadened, and it will further benefits their potential device applications in various aspects, such as correlated field emission transistor [15], fuel cells [16], thermistor [10], micro-current sensor [17], bio-sensors [18], and neuron-spin logical devices [19].

In this work, we demonstrate unexpected anisotropies in the correlated transportation properties within the electron localized insulating phase of single crystalline $Re$NiO$_3$ thin films epitaxy at various epitaxy orientations. The intrinsic and extrinsic origins associated to the anisotropic transportations of $Re$NiO$_3$ areelucidated.

From one aspect, variations in the lattice distortion when imparting bi-axial interfacial strains upon various lattice planes can result in extrinsic anisotropies in abrupt orbital transitions of $Re$NiO$_3$ and their MIT behaviors. From the other aspect, the in-plane orbital entropy ($S_{Orbit}$) associated to the in-plane symmetry of the NiO$_6$ octahedron within $Re$NiO$_3$ causes intrinsic anisotropies for the gradual orbital transition with temperature and its resultant thermistor transportations below $T_{MIT}$. It unveils the overlooked role of the electronic orbital directionality within low dimensional correlated perovskites that can trigger their anisotropic transportation behaviors, in spite of their relatively symmetric crystal structures compared to e.g. layer-structured van der Waals crystal.

**Growing $Re$NiO$_3$ thin films at various orientations:** SmNiO$_3$ (SNO), NdNiO$_3$ (NNO) and EuNiO$_3$ (ENO) thin films were epitaxially grown on the LaAlO$_3$ (LAO) and SrTiO$_3$ (STO) single crystalline substrates with various orientations of (001), (110) and (111) at a thickness of 10-20 nm, according to ref. [10]. Taking SmNiO$_3$ ($a$= 5.43 Å, $b$= 7.57 Å, $c$= 5.34 Å) as an example, the $(a^2+c^2)^{1/2}/2$ and $c/2$ of SNO (3.8 Å and 3.78 Å) is similar to the lattice constant of LaAlO$_3$ ($a_1$= $a_2$= $a_3$= 3.78 Å). As a result, the thermodynamically metastable SNO can heterogeneous nuclei and co-lattice grow on the lattice template of LaAlO$_3$ planes [10]. Figure 1a, 1b and 1c shows the corresponding cross section morphology of SNO/LAO(001), SNO/LAO(110) and SNO/LAO(111) by using high angle annular dark field (HAADF) scanning transmission electron microscope (STEM) imaging. For all three situations, SNO films are co-lattice grown on the surface of the substrate with a coherent interface in between. These observations are further supported by their X-ray diffraction (XRD) patterns shown in Figure S1 (supporting information), where diffraction peaks of thin films only appear beside the ones associated to the substrate indicating their same or equivalent crystal orientations.

Figure 1(a2-c2) and Figure 1(a3-c3) show schematic illustrations of cross-plane and in-plane lattice constructions of the $SmNiO_3$ film epitaxy on the $LaAlO_3$ substrate with the [001], [110] and [111] orientations. It can be seen that the [100] of $SmNiO_3$ is projected on the [110] projection of $LaAlO_3$ for $SmNiO_3/LaAlO_3$(001). For $SmNiO_3/LaAlO_3$(110), the [210] plane of $SmNiO_3$ is projected on the [-111] of $LaAlO_3$; while for $SmNiO_3/LaAlO_3$(111), the [001] plane of $SmNiO_3$ is projected on the [-110] of $LaAlO_3$. It is worthy to note that the in-plane construction and symmetry of the $NiO_6$ octahedron differs completely when for $SmNiO_3$ epitaxy on $LaAlO_3$ with various crystal orientations. It was pointed out that the carrier transportation of the insulating phase for rare-earth nickelates is associated to the electron hoping along the Ni-O-Ni bonds within the adjacent $NiO_6$ octahedron [9-12,20]. Therefore, an anisotropic transportation behavior is expected by confining carrier transports within the in-plane directions.

**Anisotropies in correlated transportations of $ReNiO_3$:** Figure 2a shows the temperature dependence of resistivity (*R-T*) measured for SNO/LAO(001), SNO/LAO(110) and SNO/LAO(111) from room temperature up to 180 ℃, while their $T_{MIT}$ are more clearly demonstrated by the temperature dependence of *TCR* (*TCT-T*) as shown in the inset. It clearly indicates the anisotropy in MIT when epitaxy SNO on LAO with different crystal orientations, as the $T_{MIT}$ for SNO/LAO(110) exceeds the ones for SNO/LAO(001) and SNO/LAO(111) by 15 ℃ and 10 ℃, respectively. It is worth noticing that as-observed anisotropy in MIT is not necessarily intrinsic, since the SmNiO$_3$ films co-latticed grown on LAO are biaxial compressively strained. This consideration is further confirmed, since the same $T_{MIT}$ was observed in the strain relaxed SNO/STO(001), SNO/STO(110) and SNO/STO(111), as shown in Figure S2 (supporting information).

In Figure 2b, the presently observed $T_{MIT}$ and variation in the resistivity, i.e., as evaluated by $R_{50℃}/R_{160℃}$, for SNO grown on LAO and STO at (001), (110) and (111) directions with the ones reported previously are further compared. It can be seen that the SNO/LAO(110) sample achieves the highest $T_{MIT}$ and largest $R_{50℃}/R_{160℃}$ compared to previously reported ones [10,21-24]. This result breaks the conventional understandings of $Re$NiO$_3$ that imparting biaxial compressive interfacial strain should reduce $T_{MIT}$, while in contrast the $T_{MIT}$ observed for the compressively strained SNO/LAO(110) largely exceeds the one achieved in SNO/STO(110). In addition, it is also interesting to note that despite the elevation in its $T_{MIT}$, the transition sharpness and variation in the resistivity during MIT for SNO/LAO(110) also exceed the ones achieved SNO/LAO(001) and SNO/LAO(111). This observation contradicts the conventional expectations for $Re$NiO$_3$ that the elevation in $T_{MIT}$ should reduce the transition sharpness and variation in the resistivity across MIT [10-14,20-22]. More interestingly, the $T_{MIT}$ observed for the bi-axial compressively strained SNO/LAO(110) is higher than the strain relaxed SNO/STO(110), and this also differs from the previous observations that the strained SNO/LAO(001) should exhibit a lower $T_{MIT}$ than the strain relaxed SNO/STO(001) [10,11,20,23,25]. The above results indicate the orientation-dependent orbital re-configurations via imparting asymmetric bi-axial distortions, which is more complicated than the previous understanding for SNO only grown on the STO or LAO at (001) orientations.

Further consistency observed from the temperature dependences of the thermopower for as-grown SNO/LAO(001), SNO/LAO(110) and SNO/LAO(111) is demonstrated in Figure 2c. A negative temperature dependence was observed for the thermopower with an elevated temperature, which is in agreement to their enhanced electrical conductions within the insulating phase of SNO by elevating the temperature towards the metallic phase. It is also interesting to note that a larger magnitude of thermopower was observed for the insulating phase of SNO/LAO(110), followed by SNO/LAO(111) and the smallest for SNO/LAO(001) at the same temperature. In contrast, a similar small magnitude of thermopower (25 $\mu V\ K^{-1}$) was observed for the metallic phase of these differently oriented samples.

To better investigate anisotropic transportation properties of the insulating phase of $Re$NiO$_3$, the low temperature range for SNO grown on three different oriented LAO and STO substrates are compared in Figure 2d and Figure S3 (supporting information), respectively. In general, a more significant temperature induced variation in the resistivity was observed for SNO epitaxy on the (110) oriented substrate compared to the (001) oriented one, no matter whether the interfacial strain is preserved or not, i.e. the case for LAO or relaxed, i.e. the case for STO. This is more clearly demonstrated by their respective *TCR-T* tendencies compared in Figure 2e, where a large magnitude of *TCR* over 2% $K^{-1}$ across the entire range of temperature from 100-300 K was observed for SNO/LAO(001) and SNO/STO(001). Such performance achieved in SNO epitaxy on the (110) oriented substrate benefits more for the practical thermistor applications at low temperature ranges, compared to the one achieved in other oriented samples [10]. In Figure 2f, the $\log(R)$-$T^{-1}$ relations achieved in the presently grown SNO are further compared with reported ones from SNO thin film and bulk samples. The slope in $\log(R)$-$T^{-1}$ relations indicates the *B* value ($B=-T^2TCR$) that evaluates the performance of the negative *TCR* thermistor [26], and it can be seen that the SNO/LAO(110) sample exhibits a better thermistor performance across a broad-temperature range compared to the previous reports for SNO [10,21-25].

Anisotropies in correlated transportation properties, such as the metal to insulator transition and thermistor transportation were also observed in other rare-earth nickelates, as shown in Figure S3 (supporting information) for ENO/LAO and Figure 3a for NNO/LAO. It is worthy to note that the lattice of *Re*NiO$_3$ is in-plane locked when epitaxy on the LAO substrate, as indicated by the reciprocal space mapping shown in Figure 3b and XRD patterns of NNO/LAO as show Figure S4 (supporting information, where the same in-plane lattice vectors associated to the thin film and substrate were observed. It results in biaxial in-plane compressive distortion of *Re*NiO$_3$ and the respective cross-plane transverse expansion. This is in contrast to the situation when depositing $Re$NiO$_3$ on STO, in which case the interfacial strain is completely relaxed, owning to a larger lattice mismatch between the thin film and substrate [10].

A closer comparison in the *R-T* tendencies of SNO/LAO as shown in Figure 2a or NNO/LAO as shown in Figure 3a epitaxy at various orientations indicates a more significant anisotropy observed for the insulating phase as compared to the metallic phase of *Re*NiO$_3$. The orbital structures for NNO/LAO(110), NNO/LAO(001), SNO/LAO(110) and SNO/LAO(001) at room temperature were further probed by performing the near edge X-ray absorption fine structure (NEXAF) analysis, as the results shown in Figure 3c and 3d for O-*K* edge and Ni-$L_3$ edge, respectively. In general, a more significant variation in the NEXAF spectrums between the (110) and (001) orientations were observed for SNO at the insulating phase, as compared to NNO at the metallic phase. In addition, as indicated by its larger resistivity and thermopower at the room temperature, the SNO/LAO(110) should exhibit a more strengthened insulating phase as compared to SNO/LAO(001). Nevertheless, it is interesting to note that relative intensities of the pre-peak (A) in the O-K edge and the peak (B) in the Ni-$L_3$ edge are both higher for SNO/LAO(110) compared to SNO(001), while their positions are shifted towards a slightly higher photon energy. These variations differ from tendencies observed previously when varying the rare-earth composition or imparting interfacial strains upon *Re*NiO$_3$ grown on STO or LAO at the (001) orientation [10]. It indicates the variations in the electronic structure of the insulating phase of *Re*NiO$_3$ when imparting a similar magnitude of biaxial compressive distortion along the their [010] and [001] lattice-planes, owning to the following two reasons. Firstly, $Re$NiO$_3$ exhibits a lower symmetry in the crystal structure compared to LaAlO$_3$. Secondly, the Poisson's ratio of perovskite oxides is expected to be anisotropic owning to the variations in the sheet atomic density at different orientations. In particular, an abnormal Poisson's ratio was reported for several perovskite oxides when the lattice distortion was imparted along the (110) direction [27,28].

Integrating the above demonstrations, it is worthy to note that the anisotropy achieved in the metal to insulator transition is associated to the extrinsic causes, such as anisotropic in-plane or transverse cross-plane lattice distortions. As illustrated in Figure 4a, the metal to insulator transition of $Re$NiO$_3$ is associated to the abrupt orbital transition from Ni$^{3+}$ to Ni$^{3\pm\Delta}$ as driven by the Coulomb energy [11,12,20] that is expected to be isotropic at the pristine stable (and/or metastable) state, as for the case of SNO/STO. Nevertheless, the extrinsic anisotropy is expected to be triggered via imparting bi-axial in-plane distortions upon various lattice planes that result in anisotropic lattice distortions.

In contrast, the anisotropy observed for the thermistor transportations within the insulating phase of $Re$NiO$_3$ is an intrinsic behavior, as different tendencies of $R$-$T$ (see Figure S5, supporting information) and $TCR$-$T$ (see Figure 2e) observed for the both the stained and strain-relaxed $Re$NiO$_3$ thin films. The thermistor transportation within the insulating phase of $Re$NiO$_3$ is described as $R_{(T)}=R_0\exp[E_a(k_B T)^{-1}]$, where the $E_a$ and $k_B$ represent the activation energy and Boltzmann constant, respectively [10,26]. In Figure 4c, the temperature dependences of $E_a$ are compared for SNO grown on STO and LAO substrates with orientations of (001), (110) and (111). It indicates a temperature dependence of $E_a=AT^{\alpha}+B$, where the magnitude of $\alpha$ is estimated to be 0.6, 1.9, and 0.9 for SNO/STO(001), SNO/STO(110) and SNO/STO(111), respectively.

In contrast to conventional semiconductors, the electronic transportation within the insulating phase of $Re$NiO$_3$ is associated to the carrier hopping among the NiO$_6$ octahedron [11,20]. As compared to conventional semiconductors exhibiting a relatively constant band gap (see their $R$-$T$ tendency as shown Figure S6, supporting information), the energy barrier impeding the carrier hopping among the NiO$_6$ octahedrons within the insulating phase of $Re$NiO$_3$ is temperature dependent [10]. The thicknesses of $Re$NiO$_3$ thin films grown in this work are around 10-20 nm, in which case the carrier transports are confined within the in-plane direction. As illustrated in Figure 4c, the in-plane symmetry in the two dimensional configuration of the NiO$_6$ octahedron is expected to be 4-fold, 2-fold and 3-fold, for SNO/STO(001), SNO/STO(110) and SNO/STO(111), respectively. A lower symmetry in the in-plane alignment of the nearest NiO$_6$ octahedron is expected to result in a steeper variation in the density of state (DOS) associated to the in-plane electronic conduction, as illustrated in Figure 4d. This explains a more significant variation in the resistivity when elevating (or descending) a unit fraction of energy in the form of temperature as observed for SNO at the lowest in-plane symmetry. Further imparting bi-axial compressive in-plane lattice distortion, e.g. SNO/LAO, is expected to distort the NiO$_6$ octahedron that steepens the distribution of the DOS and enlarge the magnitude of temperature coefficient of resistance, as illustrated in Figure 4e.

In summary, anisotropies in correlated transportations, such as the metal to insulator transition and thermistor transportation, are discovered in quasi-single crystalline thin films of the $d$-band correlated meta-stable rare-earth nickelates quantum systems. A significant elevation in $T_{MIT}$ was observed for $Re$NiO$_3$ co-lattice epitaxy on LaAlO$_3$ at orientations of (110) and (111) compared to (001). This originates from the extrinsic cause associated to the biaxial in-plane compressive distortion and the cross-plane transverse expansion, both of which are anisotropic. In particular, it is worthy to note that the highest $T_{MIT}$ for SmNiO$_3$ beyond ever reported was simply achieved in its thin film sample epitaxy on the (110) oriented LaAlO$_3$ substrate. In contrast, the anisotropy observed in the thermistor transportation of as-grown $Re$NiO$_3$ thin film is not only associated to the extrinsic interfacial strain, but also the intrinsic cause of the in-plane symmetry of the NiO$_6$ octahedron. Reducing the in-plane symmetry in the configuration of the nearest NiO$_6$ octahedron is expected to steepen the distribution of energy density state that elevates the temperature coefficient of resistance. As a result, the broad temperature ranged thermistor transportation properties can be significantly improved when epitaxy $Re$NiO$_3$ on the (110) oriented SrTiO$_3$ or LaAlO$_3$ substrates, compared to the ones at (001) and (111) orientations. The present demonstration enriches a new freedom in regulating electronic transportations of the electron correlated quantum system associated to the crystal orientation of the low dimensional material. It sheds a light on further enhancing electronic functionalities of correlated materials beyond the presently known merits, and this is expected to be practically useful in the field of correlated electronics.

**Acknowledgments**

This work was supported by National Natural Science Foundation of China (No. 51602022 and No. 61674013).

**Competing interests**

We declare no competing financial interest.

**Additional information:** Supplementary Information is available for this manuscript.

**Correspondence:** Prof. Jikun Chen (jikunchen@ustb.edu.cn) and Prof. Binghui Ge (bhge@ahu.edu.cn), Request for materials: Prof. Jikun Chen (jikunchen@ustb.edu.cn).

**Figures and captions**

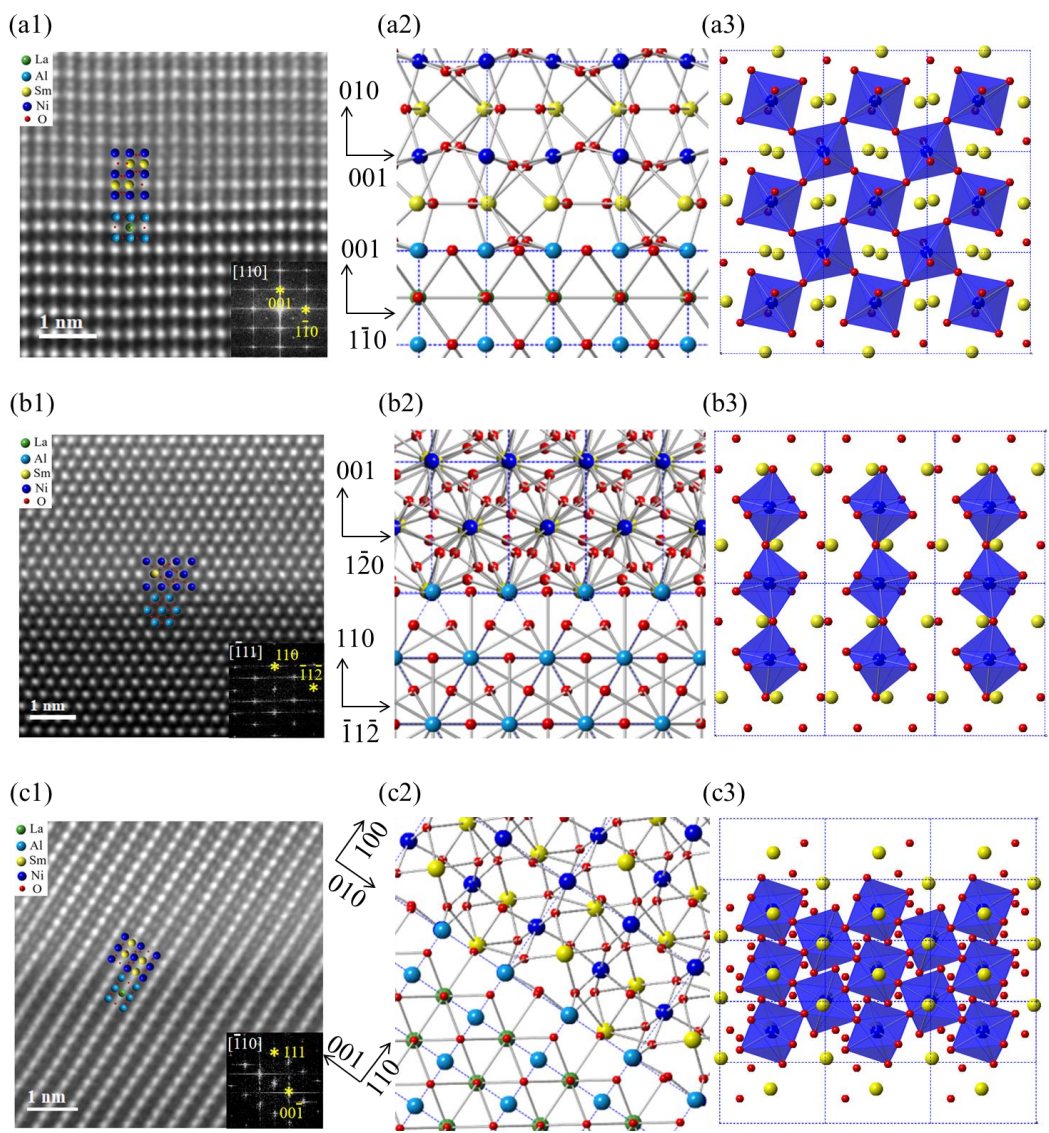

**Figure 1.** Cross-plane HAADF-STEM images **(a1)-(c1)**, cross-plane schematic models **(a2)-(c2)** and in-plane schematic models (a3-c3) of SmNiO$_3$ on LaAlO$_3$. **(a1), (a2)** [100] projection of SmNiO$_3$ on the [110] projection of LaAlO$_3$, **(b1), (b2)** [210] projection of SmNiO$_3$ on the [-111] projection of LaAlO$_3$, **(c1-c2)** [001] projection of SmNiO$_3$ on the [-110] projection of LaAlO$_3$. Inset in **(a1), (b1), (c1)** shows the fast fourier transformations of LaAlO$_3$ HAADF image. **(a3)** [010] projection of SmNiO$_3$ on the [001] oriented LaAlO$_3$ substrate, **(b3)** [001] projection of SmNiO$_3$ on the [110] oriented LaAlO$_3$ substrate, **(c3)** [210] projection of SmNiO$_3$ on the [111] oriented LaAlO$_3$ substrate.

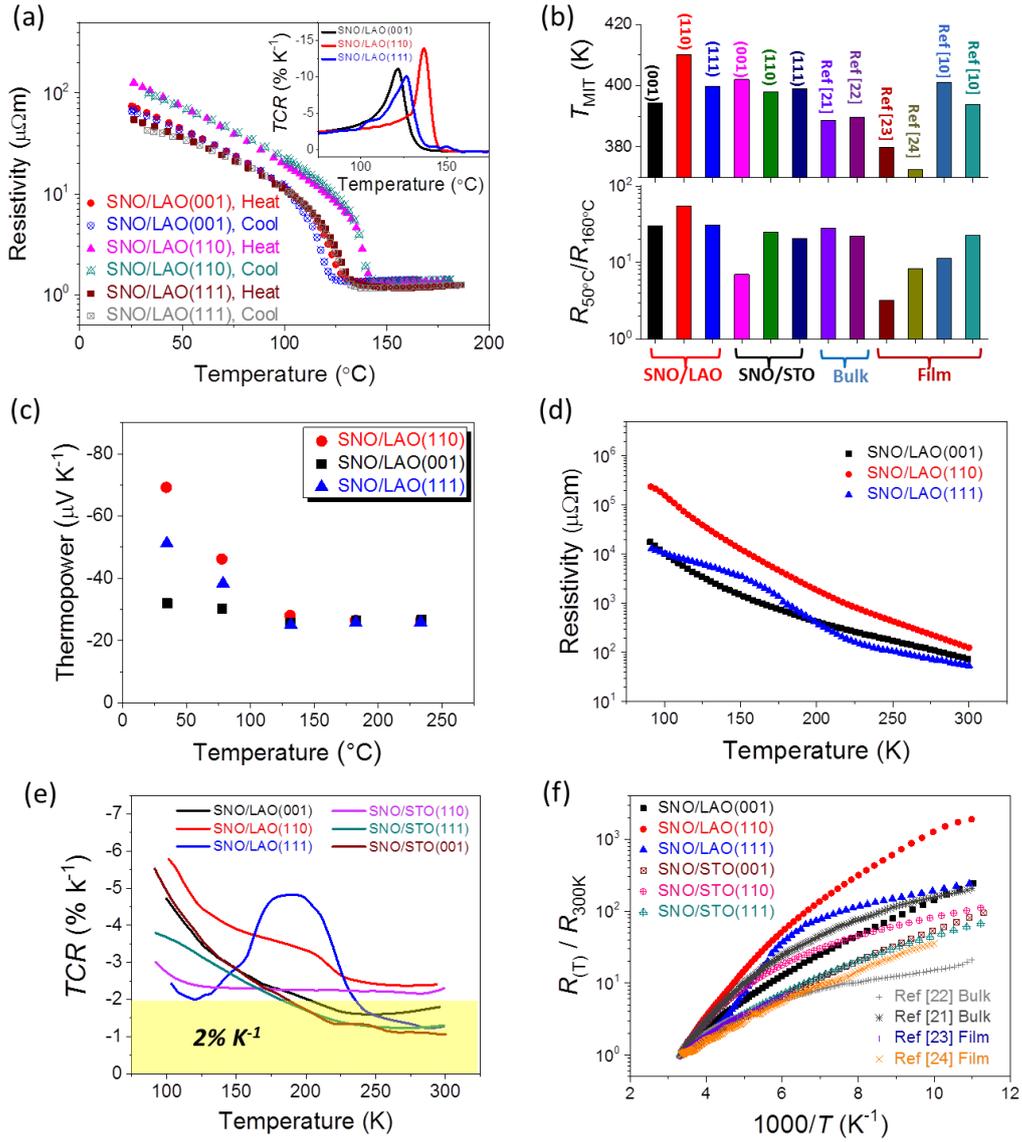

**Figure 2. (a)** Temperature-dependence of the material resistivity (*R-T*) for SmNiO$_3$ (SNO) on the single-crystalline LaAlO$_3$ (LAO) substrate with orientations of (001), (110) and (111) across their metal to insulator transition temperature ($T_{MIT}$). **(b)** Comparing the $T_{MIT}$ achieved in the SNO grown on the single crystalline LAO and SrTiO$_3$ (STO) substrates at various orientations of (001), (110) and (111) with the previous reports in refs. [11-15]. **(c)** Comparing the thermopower of SNO/LAO at (001), (110) and (111) orientations across $T_{MIT}$. **(d)** The *R-T* tendency and **(e)** respective temperature coefficient of resistance (TCR) of SNO grown on the single crystalline perovskite substrate at three orientations. **(f)** Comparing the log(*R*)-$T^{-1}$ tendency achieved in SNO/LAO and SNO/STO at three orientations in this work with previously reported ones from refs. [12-16].

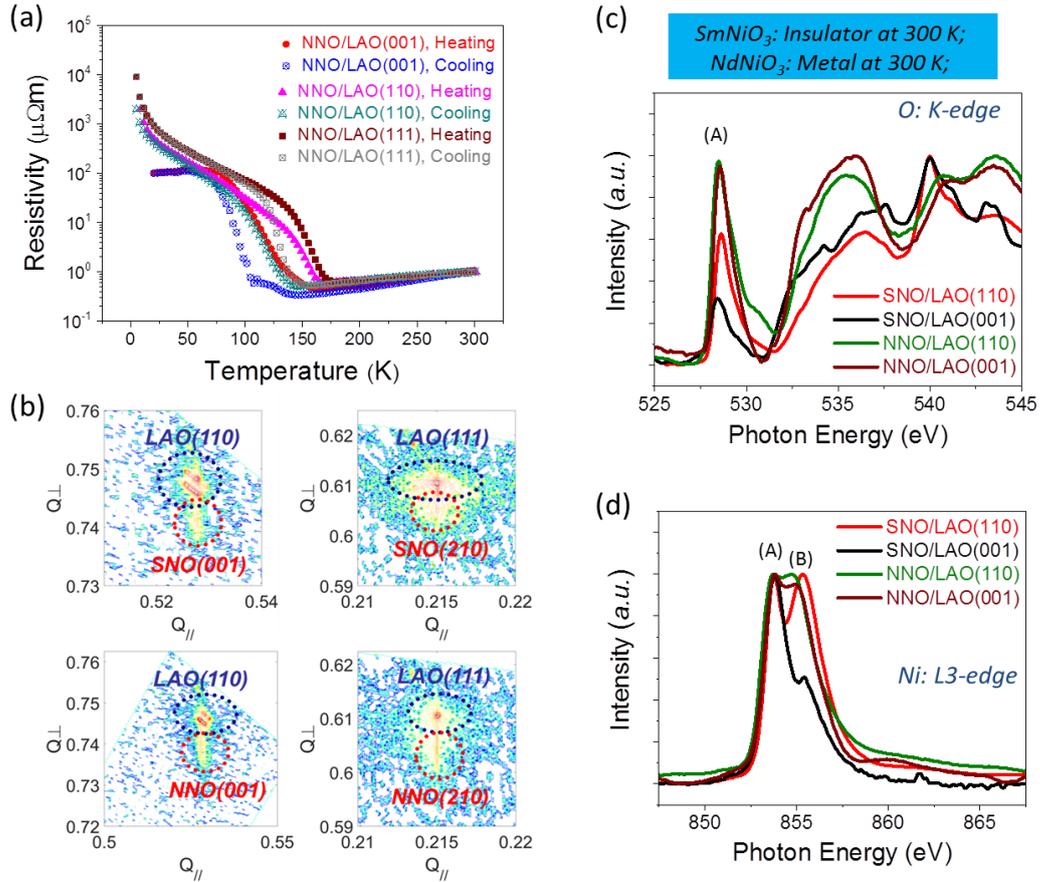

**Figure 3.** **(a)** Temperature-dependence of the material resistivity (*R-T*) for NNO/LAO at orientations of (001), (110) and (111) across their metal to insulator transition temperature ($T_{MIT}$). **(b)** The reciprocal space mapping (RSM) of SmNiO$_3$ (SNO) and NdNiO$_3$ (NNO) grown on the single crystalline LaAlO$_3$ (LAO) substrate at (110) and (111) orientations. **(c), (d)** Comparing the near edge X-ray absorption fine structure (NEXAFS) analysis of **(c)** O-*K* edge and **(d)** Ni-*L$_3$* edge for SmNiO$_3$ (SNO) and NNO grown on LAO(001) and LAO(110).

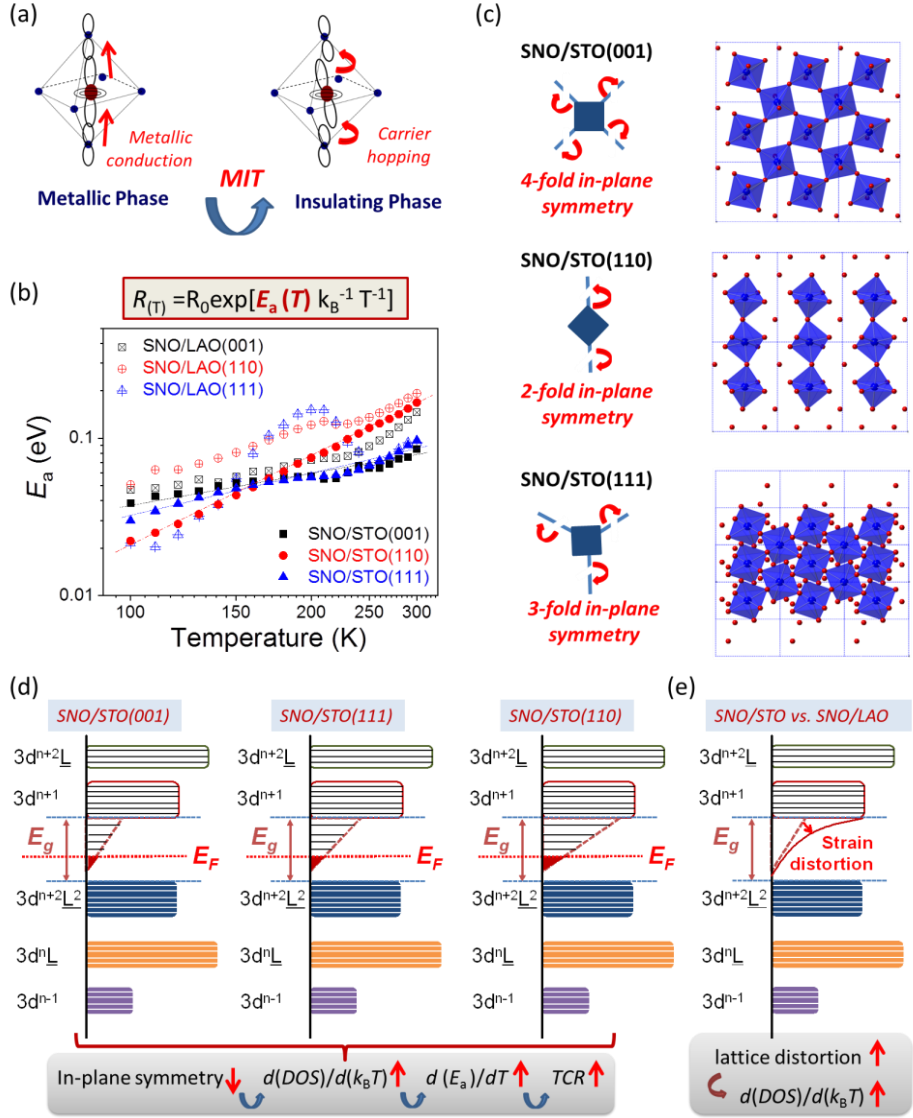

**Figure 4.** **(a)** Illustrating the orbital transition when $Re$NiO$_3$ experience the metal to insulator transition. **(b)** Activation energy ($E_a$) associated to the thermistor transportation of the bi-axial compressive stained (SNO/LAO) and unstrained (SNO/STO) SmNiO$_3$ samples epitaxy at various orientations. **(c)** Illustrating the in-plane symmetry in the alignment of the adjacent NiO$_6$ octahedron and carrier transportations for SmNiO$_3$ films epitaxy on SrTiO$_3$ substrates at various orientations. A 2-fold in-plane symmetry (2-fold) is achieved for the case of epitaxy on STO(110), 3-fold for STO(111), and 4 fold for STO(001). **(d)** Confining the carrier hopping within the in-plane direction and elevating the same fraction of phonon energy, the coupling in the orbital density of states (DOS) is expected to be the most sensitive for SNO/STO(110), followed by SNO/STO(111) and the least for SNO/STO(001). **(e)** Further imparting bi-axial compressive in-plane lattice distortion is expected to distort the NiO$_6$ octahedron that steepens the distribution of the DOS and enlarge the magnitude of temperature coefficient of resistance

# References

[1] Ben fez, L. A.; Sierra, J. F.; Torres, W. S.; Arrighi, A.; Bonell, F.; Costache, M. V.; and Valenzue, S. O. Strongly anisotropic spin relaxation in graphene-transition metal dichalcogenide heterostructures at room temperature. *Nat. Phys.* **2018**, 14, 303–308

[2] Gong, C.; Li, L.; Li, Z.; Ji, H.; Stern, A.; Xia, Y.; Cao, T.; Bao, W.; Wang, C.; Wang, Y.; Qiu, Z. Q.; Cava, R. J.; Louie, S. G.; Xia, J.; Zhang, X. Discovery of intrinsic ferromagnetism in two-dimensional van der Waals crystals. *Nature*, **2017**, 546, 265

[3] Huang, B.; Clark, G.; Navarro-Moratalla, E.; Klein, D. R.; Cheng, R.; Seyler, K. L.; Zhong, D.; Schmidgall, E.; McGuire, M. A.; Cobden, D. H.; Yao, W.; Xiao, D.; Jarillo-Herrero, P.; Xu, X. Layer-dependent ferromagnetism in a van der Waals crystal down to the monolayer limit. *Nature*, **2017**, 546, 270

[4] Fanchiang, Y. T.; Chen, K. H. M.; Tseng, C. C.; Chen, C. C.; Cheng, C. K.; Yang, S. R.; Wu, C. N.; Lee, S. F.; Hong, M.; Kwo, J. Strongly exchange-coupled and surface-statemodulated magnetization dynamics in Bi2Se3/yttrium iron garnet heterostructures. *Nat. Commun.* **2018**, 9, 223

[5] Kim, J.; Kim, K. W.; Wang, H.; Sinova, J.; Wu, R. Understanding the giant enhancement of exchange interaction in Bi2Se3-EuS heterostructures. *Phys. Rev. Lett.* **2017**, 119, 027201

[6] Han, J.; Richardella, A.; Siddiqui, S. A.; Finley, J.; Samarth, N.; Liu, L. Room-temperature spin-orbit torque switching induced by a topological insulator. *Phys. Rev. Lett.* **2017**, 119, 077702

[7] Mott, N. F. Metal-insulator transition. *Rev. Mod. Phys.*, **1968**, 40, 677-783

[8] Dagotto, E.; Correlated electrons in high-temperature superconductors. *Rev. Mod. Phys.*, **1994**, 66, 763

[9] R. Jaramillo, S. D. Ha, D. M. Silevitch, S. Ramanathan, Origins of bad-metal conductivity and the insulator–metal transition in the rare-earth nickelates. *Nat. Phys.* **2014**, 10, 304

[10] Chen, J.; Hu, H.; Wang, J.; Yajima, T.; Ge, B.; Ke, X.; Dong, H.; Jiang, Y.; Chen, N., Overcoming synthetic metastabilities and revealing metal-to-insulator transition & thermistor bi-functionalities for d-band correlation perovskite nickelates. *Mater. Horiz.* **2019**, 6, 788

[11] G. Catalan, Progress in perovskite nickelate research. *Phase Trans.* **2008**, 81, 729.

[12] I. I. Mazin, D. I. Khomskii, R. Lengsdorf, J. A. Alonso, W. G. Marshall, R. M. Ibberson, A. Podlesnyak, M. J. Martíñez-Lope, and M. M. Abd-Elmeguid, Charge Ordering as Alternative to Jahn-Teller Distortion. *Phys. Rev. Lett.* **2007**, 98, 176406

[13] J. S. Zhou and J. B. Goodenough, Chemical bonding and electronic structure of $R$NiO$_3$

(*R*=rare earth), *Phys. Rev. B* **2004**, 69, 153105

[14] J. A. Alonso, J. L. García-Muñoz, M. T. Fernández-Díaz, M. A. G. Aranda, M. J. Martínez-Lope, and M. T. Casais, Charge Disproportionation in *R*NiO$_3$ Perovskites: Simultaneous metal-insulator and structural transition in YNiO$_3$. *Phys. Rev. Lett.* **1999**, 82, 3871

[15] Shi, J.; Ha, S. D.; Zhou, Y.; Schoofs, F.; Ramanathan, S. A correlated nickelate synaptic transistor. *Nat. Commun.* **2013**, 4, 2676

[16] Zhou, Y.; Guan, X.; Zhou, H.; Ramadoss, K.; Adam, S.; Liu, H.; Lee, S.; Shi, J.; Tsuchiya, M.; Fong D. D.; Ramanathan, S. Strongly correlated perovskite fuel cells. *Nature* **2016**, 534, 231

[17] Zhang, Z.; Schwanz, D.; Narayanan, B.; Kotiuga, M.; Dura, J. A.; Cherukara, M.; Zhou, H.; Freeland, J. W.; Li, J.; Sutarto, R.; He, F.; Wu, C.; Zhu, J.; Sun, Y.; Ramadoss, K.; Nonnenmann, S. S.; Yu, N.; Comin, R.; Rabe, K. M.; Sankaranarayanan, S. K. R. S.; Ramanathan, S. Rerovskite nickelates as electric-field sensor in salt water. *Nature* **2018**, 553, 68

[18] Zhang, H. T.; Zuo, F.; Li, F.; Chan, H.; Wu, Q.; Zhang, Z.; Narayanan, B.; Ramadoss, K.; Chakraborty, I.; Saha, G.; Kamath, G.; Roy, K.; Zhou, H.; Chubykin, A. A.; Sankaranarayanan, S. K. R. S.; Choi, J. H.; Ramanathan, S. Perovskite nickelates as bio-electronic interfaces. *Nat. Commun.* **2019**, 10, 1651

[19] Zuo, F.; Panda, P.; Kotiuga, M.; Li, J.; Kang, M.; Mazzoli, C.; Zhou, H.; Barbour, A.; Wilkins, S.; Narayanan, B.; Cherukara, M.; Zhang, Z.; Sankaranarayanan, S. K. R. S.; Comin, R.; Rabe, K. M.; Roy K.; Ramanathan S. Habituation based synaptic plasticity and organismic learning in a quantum perovskite. *Nat. Commun.* **2017**, 8, 240

[20] Varignon, J.; Grisolia1, M. N.; Íñiguez, J.; Barthélémy, A.; Bibes, M. Complete phase diagram of rare-earth nickelates from firstprinciples. *NPJ Quant. Mater.* **2017**, 2, 21

[21] I. V. Nikulina, M. A. Novojilova, A. R. Kaulb, S. N. Mudretsovab and S. V. Kondrashovb, Oxygen nonstoichiometry of NdNiO$_3$ and SmNiO$_3$. *Mater. Res. Bull.,* 2004, **39**, 775–791

[22] M. T. Escote, A. M. L. da Silva, J. R. Matos and R. F. Jardim, General properties of polycrystalline *Ln*NiO$_3$ (Ln = Pr, Nd, Sm) compounds prepared through different precursors. *J. Solid State Chem.* 2000, **151**, 298-307

[23] F. Conchon, A. Boulle and R. Guinebretière, Effect of tensile and compressive strains on the transport properties of SmNiO$_3$ layers epitaxially grown on (001) SrTiO$_3$ and LaAlO$_3$ substrates. *Appl. Phys. Lett.,* 2007, **91**, 192110.

[24] Shukla, N.; Joshi, T.; Dasgupta, S.; Borisov, P. Lederman, D.; Datta S. Electrically induced insulator to metal transition in epitaxial SmNiO$_3$ thin films. *Appl. Phys. Lett.* **2014**, 105, 012108

[25] Catalano, S.; Gibert, M.; Bisogni, V.; Peil1, O. E.; He, F.; Sutarto, R.; Viret, M.; Zubko, P.; Scherwitzl, R.; Georges, A.; Sawatzky, G. A.; Schmitt, T.; Triscone, J. M. Electronic transitions in strained SmNiO3 thin films. *APL Mater.* **2014**, 2, 116110

[26] Ohg, E. Negative temperature coefficient resistance (NTCR) ceramic thermistors: An industrial perspective. *J. Am. Ceram. Soc.,* **2009**, 92, 967–983.

[27] Huang, C. W.; Ren, W.; Nguyen, V. C.; Chen, Z.; Wang, J.; Sritharan, T.; Chen, L. Abnormal Poisson's ratio and linear compressibility in perovskite materials. *Adv. Mater.* **2012**, 24, 4170–4174

[28] Chen, S.; Guan, C.; Ke, S.; Zeng, X.; Huang, C. Hu, S.; Yen, F.; Huang, H.; Lu, Y.; Chen, L. Modulation of Abnormal Poisson's ratios and electronic properties in mixed-valence perovskite manganite films. *ACS Appl. Mater. Inter.* **2018**, 10, 18029−18035

# Supporting Information

# Overlooked transportation anisotropies in *d*-band correlated rare-earth perovskite nickelates

*Jikun Chen[1]\*, Haiyang Hu[1], Fanqi Meng[2], Takeaki Yajima[3], Lixia Yang[2], Binghui Ge[2,4]\*, Xinyou Ke[5], Jiaou Wang[6], Yong Jiang[1], Nuofu Chen[7]*

[1]Beijing Advanced Innovation Center for Materials Genome Engineering, School of Materials Science and Engineering, University of Science and Technology Beijing, Beijing 100083, China
[2]Beijing National Laboratory for Condensed Matter Physics, Institute of Physics, Chinese Academy of Sciences, 100190 Beijing, China
[3]School of Engineering, The University of Tokyo, 2-11-16 Yayoi, Bunkyo-ku, Tokyo 113-0032, Japan
[4]Institutes of Physical Science and Information Technology, Anhui University, Hefei, Anhui 230601, China
[5]Department of Mechanical and Aerospace Engineering, Case Western Reserve University, Cleveland, Ohio 44106, USA
[6]Beijing Synchrotron Radiation Facility, Institute of High Energy Physics, Chinese Academy of Sciences, Beijing 100049, China
[7]School of Renewable Energy, North China Electric Power University, Beijing 102206, China

Correspondence: Prof. Jikun Chen (jikunchen@ustb.edu.cn) and Prof. Binghui Ge (bhge@ahu.edu.cn)

**Section A: Experimental details**

**Sample growth:** The thin films of $Re$NiO$_3$, including SmNiO$_3$ (SNO), NdNiO$_3$ (NNO), and EuNiO$_3$ (ENO), were grown on substrates of the single crystalline perovskites, such as LaAlO$_3$ (LAO), SrTiO$_3$ (STO) and (LaAlO$_3$)$_{0.3}$(Sr$_2$AlTaO$_6$)$_{0.7}$ (LSAT) at various orientations of (001), (110) and (111). The thin film growth strategy was previously reported in our previous work [10]. In brief, chemical precursors of $Re$(NO$_3$)$_3$ and Ni(CH$_3$COOH)$_2$ were mixed at stoichiometric ratio, and dissolved in ethylene glycol monomethyl ether (EGME). The chemical solution was spin coated on the surface of perovskite substrates, dried and annealed at high oxygen pressures from 15 to 20 MPa, and 800 °C.

**Characterizations:** The temperature dependences of the resistivity (*R-T*) of as-grown thin films were measured by a CTA-system in the temperature range above room temperature, while the *R-T* tendency below room temperature was characterized by using a PPMS system (Quantum Design). The crystal structure along the cross-plane direction was measured by a conventional *θ*-2*θ* scan of X-ray diffractions, while the in-plane lattice structure was measured by reciprocal space mapping (RSM). The cross-section morphologies of the films were characterized by high-angle annular dark-field (HAADF) and annular bright-field (ABF) scanning transmission electron microscopy (STEM), on JEM-ARM 200F TEM operated at 200 kV with a cold field emission gun and aberration correctors for both probe-forming and imaging lenses. The near edge X-ray absorption fine structure (NEXAFS) at Beijing Synchrotron Radiation Facility, Institute of High Energy Physics, Chinese Academy of Sciences, Beijing 100049, China.

**Section B: Additional results**

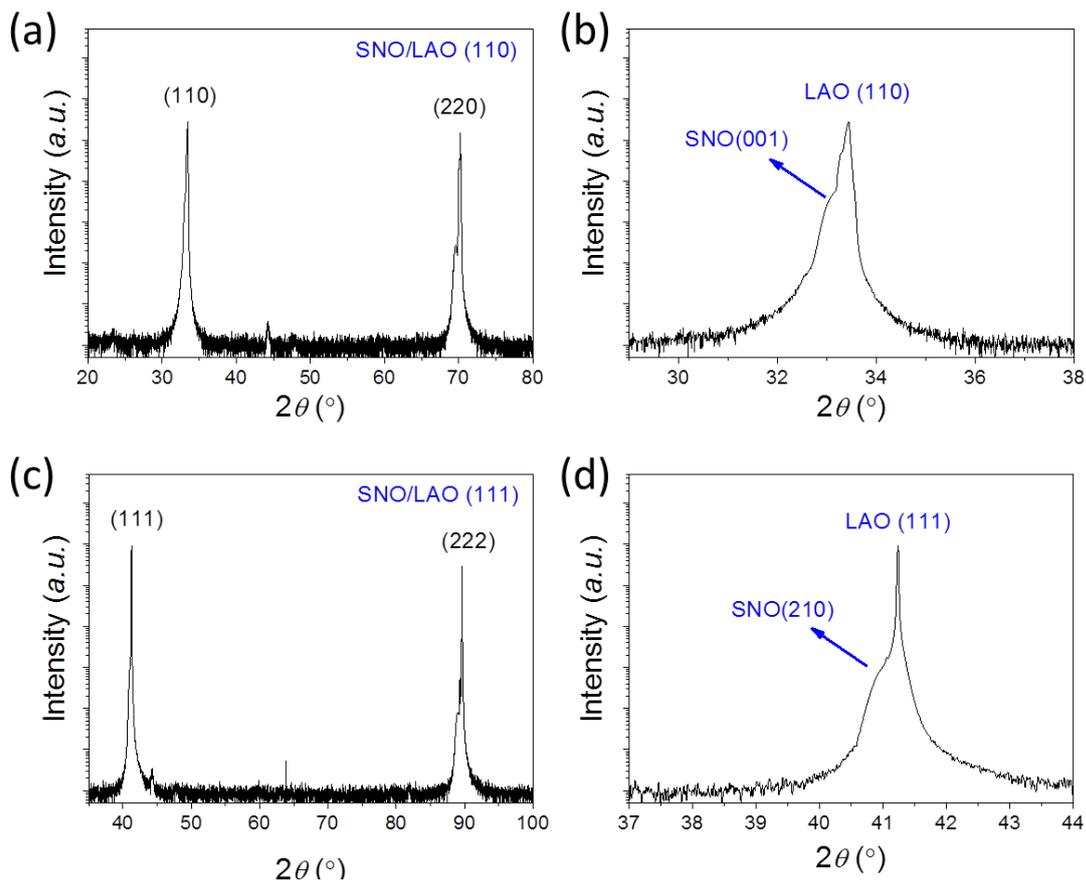

**Figure S1.** X-ray diffraction patterns of SmNiO$_3$ (SNO) grown on the single crystalline LaAlO$_3$ (LAO) substrates at orientations of **(a), (b)** (110) and **(c), (d)** (111).

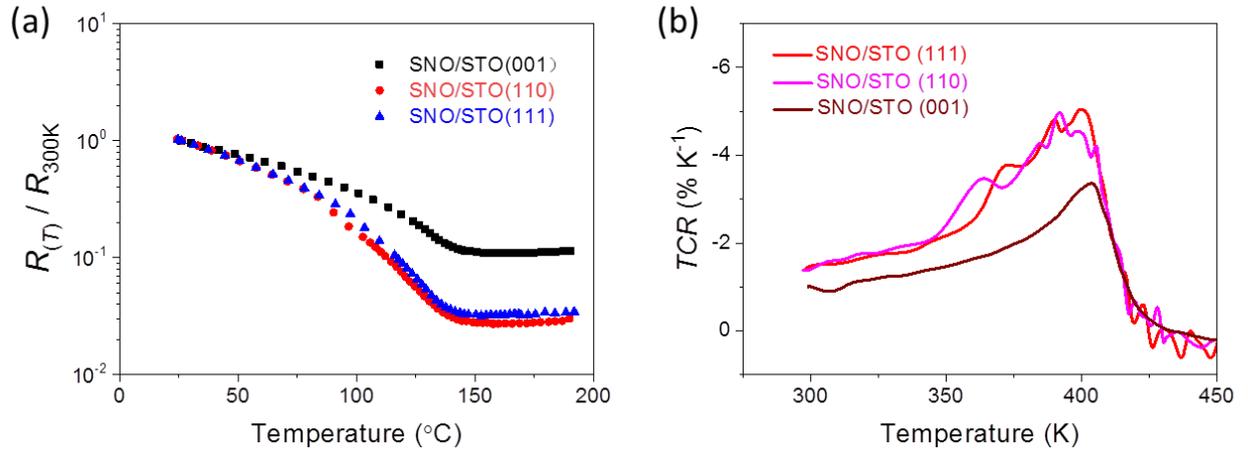

**Figure S2.** The temperature dependence of **(a)** resistivity and **(b)** temperature coefficient of resistance (TCR) for as grown $SmNiO_3$ on the single crystalline $SrTiO_3$ substrates at (001), (110) and (111) orientations. The same metal to insulator transition temperature was observed for three different epitaxy orientations.

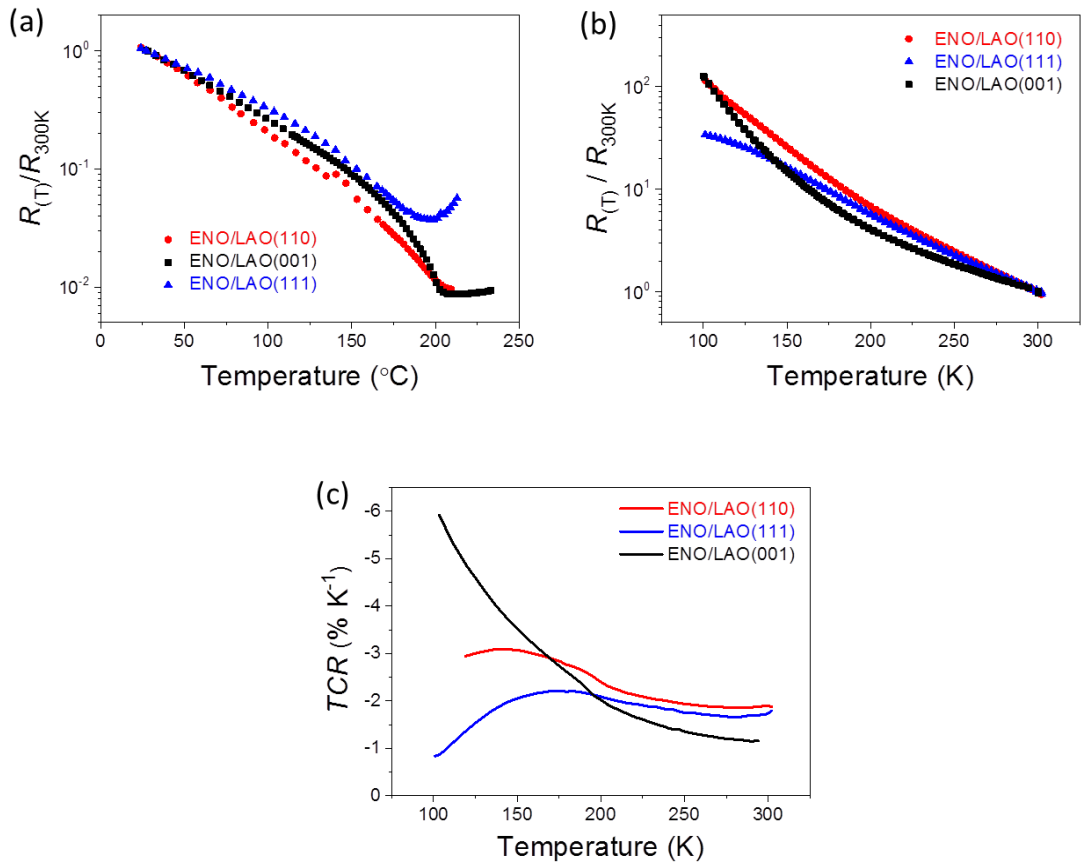

**Figure S3.** The temperature dependence of **(a), (b)** resistivity above and below room temperature, and **(c)** temperature coefficient of resistance (TCR) for as grown $EuNiO_3$ on the single crystalline $LaAlO_3$ substrates at (001), (110) and (111) orientations. Both the temperature dependence of resistivity and TCR were observed.

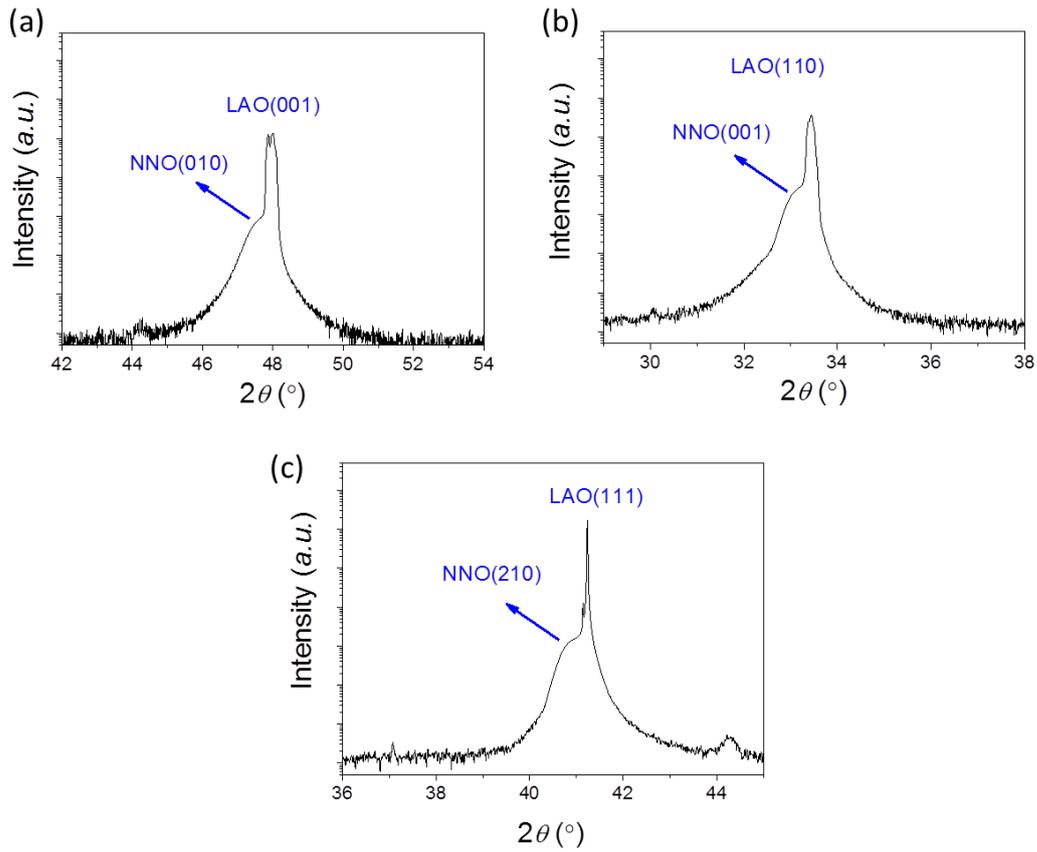

**Figure S4.** X-ray diffraction patterns of NdNiO$_3$ (NNO) grown on the single crystalline LaAlO$_3$ (LAO) substrates at orientations of **(a)** (001) **(b)** (110), and **(c)** (111).

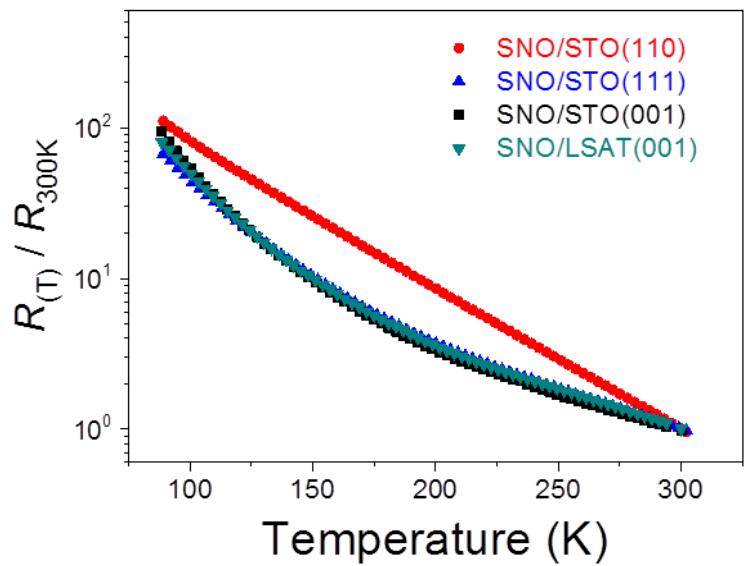

**Figure S5.** The temperature dependence of the strain relaxed SmNiO$_3$ grown on the single crystalline SrTiO$_3$ substrates at (001), (110) and (111) orientations as well as the LSAT(001) substrate.

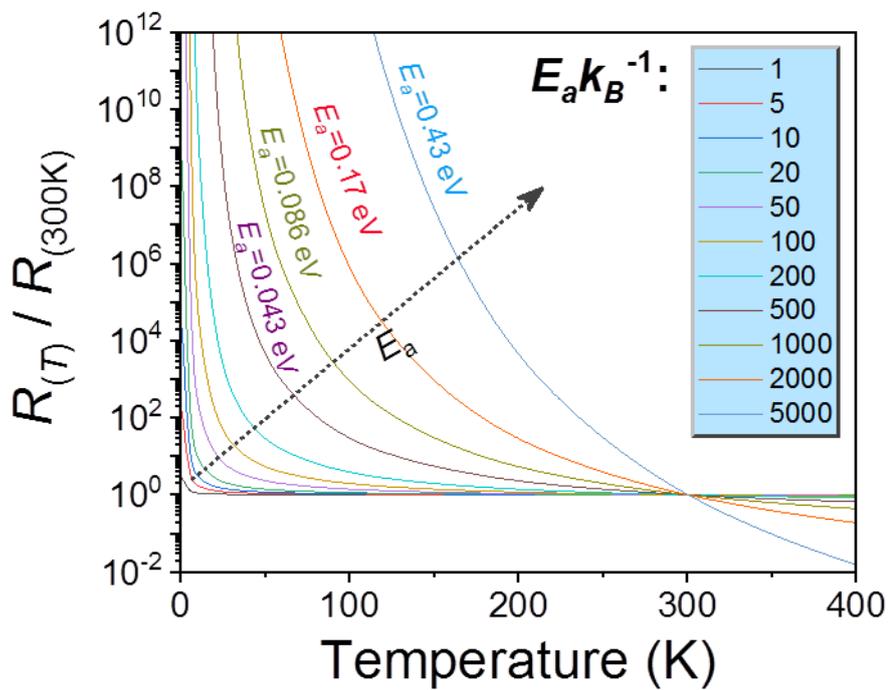

**Figure S6.** The temperature dependence in resistivity (*R-T*) for negative temperature coefficient of resistance thermistor transportations at constant magnitudes of $E_a$.